# ViaPPS: A Mobile Pavement Profiling System

*Henri Giudici[1], Boris Mocialov[2], Aslak Myklatun[3]*

**ViaTech AS**

Ensuring safety levels on roads is an imperative task for road authorities. Significant amounts of time and money are spent on performing road inspections every year. In order to ensure efficiency of road inventories, road practitioners are in need of reliable systems that are fast and are designed to comply with high standards at lower costs. To date, the 3D Mobile Mapping Systems is the most efficient, productive and accurate system used during evaluation of pavement inventories. This paper presents ViaPPS – a 3D Mobile Mapping System from ViaTech AS. ViaPPS offers accurate georeferenced data by combining the perception and navigation sensors. This paper showcases results from multiple mobile systems after the harmonization process which is held once a year to evaluate reliability and repeatability of the measurements.

1: Ph.D. Principal Scientist: henri@viatch.no

2: Ph.D. Computer Vision Engineer

3: Head of Research and Development

## 1. Introduction

3D Mobile Mapping Systems are widely used for pavement inventories due to their highly accurate georeferenced sensing abilities [1]. Those systems are composed of multiple mobile sensors divided into perception and position/navigation sensors [2]. Some of the examples of the perception sensors are the laser scanners, laser texture scanners (often called profilometers), RGB cameras while the Inertial Measurement Unit (IMU), Global Position Systems (GPS), and odometer are members of the navigation-position sensors.

The presented laser scanners are based on LIDAR technology and the laser profilometers are based on laser triangulation technology. Both laser scanners and laser profilometers are widely adopted in various industries for their ability to capture details of the surroundings in point cloud form (laser scanners) and in point by point form (laser profilometers) [3,4,7]. Laser profilometers are often used to investigate the roughness of surfaces and factors that affect the vehicle pavement interaction at micro, macro and mega scale (e.g. tire-friction, tire-wear, hydroplaning, rolling resistance, etc) [5,6,7].

The laser scanners point cloud is adopted for the evaluation of pavement condition such as: road defects (cracks, potholes, ravelling, etc), road markings, pavement cross fall and many others. The laser profilometers measurements are used for the pavement texture description with the use of two parameters: Mean Profile Depth (MPD), for the characterization of texture's depth and International Roughness Index (IRI) for the longitudinal road profile roughness.

The IMU combines gyro and accelerometers generating accurate motion information of a 3D system which can be adopted in various industries [8]. GPS receiver provides position information from Global Navigation Satellite System (GNSS). Counting the wheel rotations, the odometer sensor measures the vehicle travelled distance. The fusion of IMU, GPS and odometer sensors captures the motion of a vehicle in various environments including tunnels or sub-urban areas. In case of missing GPS signals, the vehicle position is estimated with the combination of IMU and odometer information.

This paper describes ViaPPS - a 3D Mobile Mapping System from ViaTech AS. ViaPPS is a multipurpose mobile system designed for, but not limited to, road inventories, complying with different international standards which are listed in references: [9-19]. ViaPPS consists of multiple light-weight portable sensors mounted on a vehicle as base platform. The multiple sensors are: 3 laser scanners, laser profilometer, an IMU, GPS, 4 RGB cameras, and odometer mounted on wheel rim (see Figure 1). To check the reliability and repeatability of the systems measurements, each year the systems gather in a testing area located in Fredrikstad (Norway) in good weather condition for the harmonization process. During the harmonization process the systems measurements are compared against each other in order to verify that the deviation between systems measurements is within acceptable low threshold levels. Further, this paper will describe the harmonization results.

## 2. ViaPPS

*2.1 System Design*

The 3 laser scanners are mounted at the back upper side of the platform (1). At the top of (1) a 360° camera is located (2), while at the bottom of (1) an inclined camera points at the road surface (3). The IMU is inside the vehicle (4). A laser profilometer (5) is mounted on the front part of the vehicle, next to the right wheel, while an odometer is located next to the front wheel rim (6). The two GPS antennas (7) are in front of the vehicle and next to (2). The PC is next to the operator, controlling the whole system (8). Two cameras are mounted in front of the vehicle next to the operator position (9). Table 1 shows the mobile sensors on the vehicle platform.

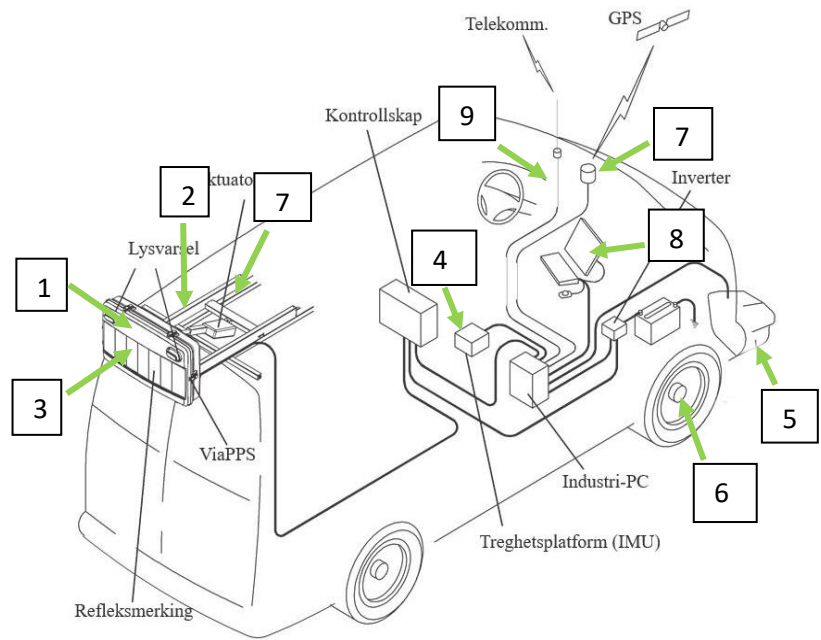

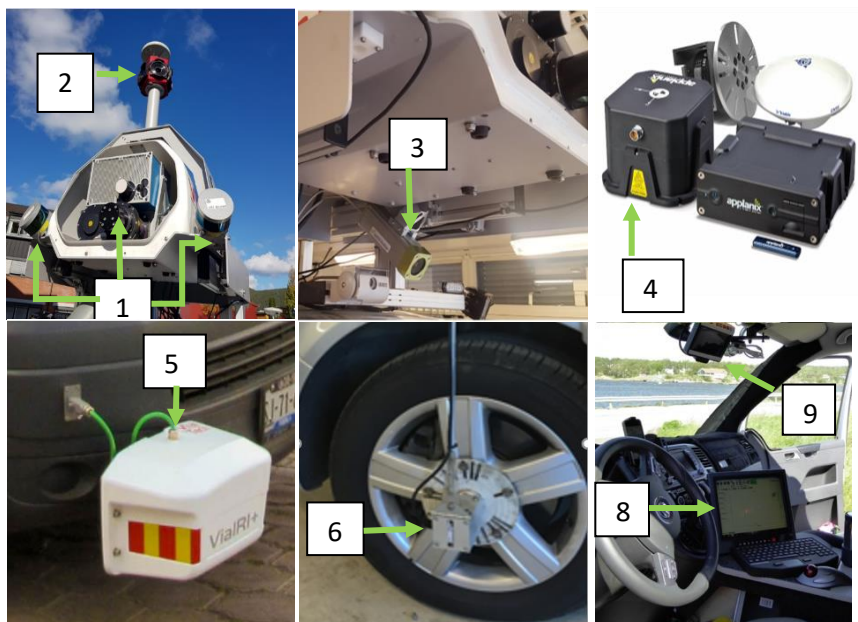

**Figure 1** Individual components of the ViaPPS system.

**Table 1** ViaPPS mobile sensors.

| Sensor | Model | Characteristics |
|---|---|---|
| Laser scanner | Z+F Profiler 9012A / | - Phased based technology<br>360 ° field view<br>- Protection class IP54<br>- 200 profiles/sec<br>- Over 1 x 10$^6$ pps (points pr second)<br>- -10 °C < working temp. < 45 ° C<br>- Accuracy < 1mm |
| Laser scanner | 2 Velodyne Puck Lite<br>Supports any laser<br>scanner from Velodyne | Time of flight technology<br>360 ° FOV<br>330.000 pps<br>2 – 3 cm accuracy |
| Laser profilometer (ViaIRI) | Riftek | - Laser safety class: L – Class 2<br>- Shock absorbent<br>- -10° C < working temp. < 60° C |
| Planar Camera System | Cameras from Basler | Maximum 8 cameras<br>5 MP MPX<br>Image recording by configurable<br>distance interval [m] or time interval. |
| 360 Camera System | Ladybug 5+ from FLIR | Covers 90% of the sphere<br>6 cameras, one pointing upwards<br>5 MPX resolution<br>15 FMPS<br>Image recording by configurable<br>distance interval [m] or time interval. |
| Navigation System | IMU - Applanix POS LV<br>GPS – Trimble<br>Odometer – Applanix | - GNSS - integrated inertial technology<br>- Solid State MEMS IMU<br>- Support Distance Measurement Indicator<br>- GPS Receivers and Antennas |

## 2.2 Data processing

### Calibration Perception – Position/navigation sensors

To provide accurate integration of their relative data, perception and position/navigation sensors need a common reference system [2]. For this reason, the sensors are adjusted to a common reference system. As result each of the perception sensors data (i.e. single scan and/or image) provide accurate georeferenced information.

*Calibration RGB camera – LIDAR sensors*

Calibration between an RGB camera and a LiDAR sensor requires estimation of extrinsic and intrinsic parameters of the camera. While linear intrinsic parameters contain information about the focal length ($f$) and the principal point ($c$) and are encoded in a matrix $K$, non-linear intrinsic parameters can describe lens distortion coefficients ($k_n$). Similarly, extrinsic parameters describe the rotation ($R$) and translation ($t$) between the two sensors. Putting the individual parts together gives a model that can describe a conceptual pinhole camera $[R/t]K$, where $K = \begin{pmatrix} f_x & 0 & c_x \\ 0 & f_y & c_y \\ 0 & 0 & 1 \end{pmatrix}$. All the sensors are calibrated to a fixed common vehicle reference point and adjusted with the GNSS system as well as synchronized in time. Consequently, each sensor (scan or image) contain the same timestamp and geo location as follows: Time, fixed reference system (X, Y, Z), fixed rotational angles (Yaw, Pitch, Roll). In order to model a real camera that contains a lens, distortion parameters should be added to the equation

$(r^2 k_1 + r^4 k_2 + r^6 k_3 + \cdots)$, where r is the distance from a point (x, y) to the center of radial distortion defined as $r = \sqrt{x^2 + y^2}$ [25]. It is common that the intrinsic parameters ($f, c, k_n$) of a camera are not known a priori, which makes it difficult to estimate the camera matrix [24]. Luckily, optimisation algorithms can be used to estimate the intrinsic and/or extrinsic parameters, such as RANSAC for the calibration of extrinsic parameters [20, 21] Levenberg-Marquardt for refinement of the closed-form solution [22], and machine learning for estimating the intrinsic and extrinsic transformations of LiDAR to camera [23].

We use the direct linear transform (DLT) [27] optimisation algorithm to find a projection matrix (homography) from a known set of point mappings. First, we normalise the data by translating by the mean and scaling by the standard deviation to make the algorithm less sensitive to the outliers and lens distortion. Then, we turn the homography equation $\begin{bmatrix} x' \\ y' \\ 1 \end{bmatrix} = \alpha H \begin{bmatrix} x \\ y \\ z \end{bmatrix}$ into homogeneous linear equations. Finally, we solve the homogeneous system using the Singular-Value Decomposition method and use the eigenvector corresponding to the smallest eigenvalue and perform de-normalisation to obtain the projection matrix H. The problem with using the DLT method alone is that it ignores the distortion parameters ($k_n$) that are introduced by the camera lens. Another problem with the DLT method is that it estimates the homography, but does not directly provide the individual camera parameters [25]. In addition to the DLT optimisation, we use the Levenberg-Marquardt (LM) [28, 29] method in order to minimise the projection error introduced by the noise in the data used for the calibration. The cost of the LM algorithm is the reprojection error $(x_i' - x_i)^2 + (y_i' - y_i)^2$ between the projected (x', y') and observed (x, y) points that is minimised using the gradient descent method [26].

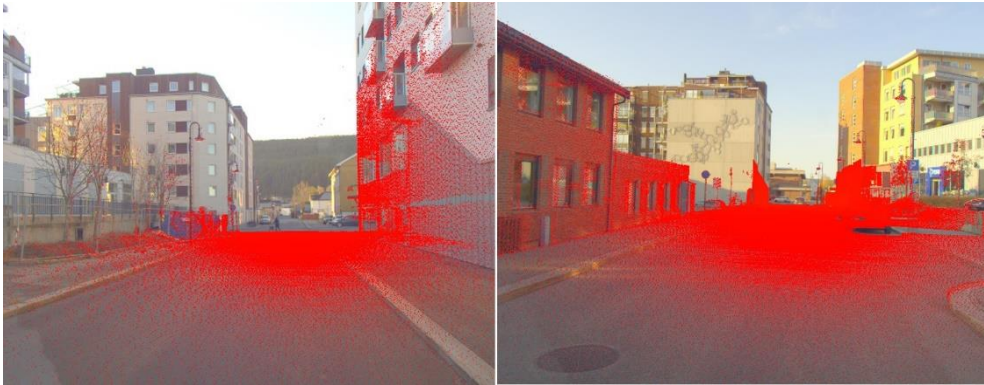

**Figure 2** RGB image from the VPS system with the corresponding LiDAR points from the Velodyne sensor.

Obtained 3D point cloud is used to detect the presence of various defects in the pavement with the help of a road distress adaptive algorithm. For example: Figure 3 shows the detection of a cracks on road while Figure 4 shows the detection of road edges and road markings.

## 2.3 System operation

The operator activates the system using a graphical user interface. Once the system has been activated, the operator inspects the pavement by driving on its surface and collecting the data using all the sensors available on the ViaPPS system. At the end of the inspection, the operator terminates the data collection. The collected data is then analysed, and a report is generated. Figures 3 and 4 show the interface with a generated report. The data analysis includes discovery of interesting features for pavement inventory (e.g. cracks, potholes, ravelling, etc).

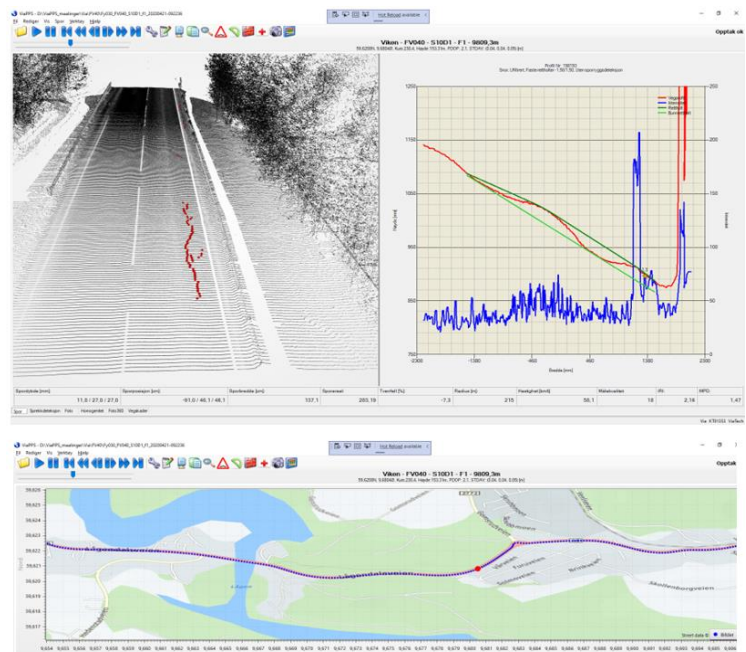

**Figure 3** a) 3D pavement point cloud with crack detection; b) pavement profile from single scan; c) positions of the inspected pavement.

Figure 3a shows the resulting 3D point cloud of the road surface with a detected crack (in red), figure 3b shows a single profile scan (in red) together with the relative light intensity of the scan (in blue). Figure 3c shows the reported latitude and longitude mapped data while operating the system. Figures 4a and b show the detection of road edges and road markings from point clouds.

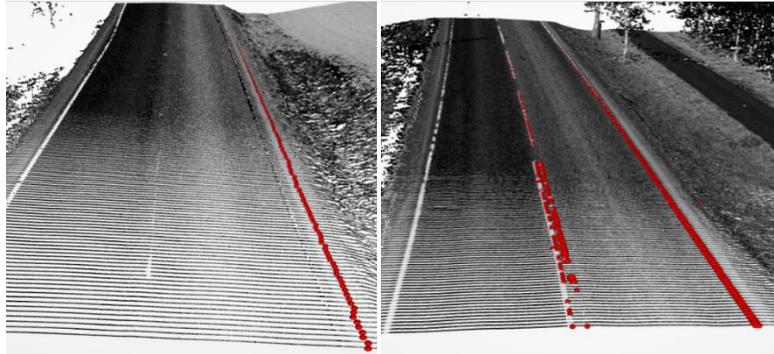

**Figure 4** 3D point cloud with: a) road edge detection; b) road marking detection.

## 3 System reliability: Harmonization process

To check the reliability and repeatability between the different 3D Mobile Mapping Systems, every year the systems gather together in a specific testing area in Norway for the harmonization process. During the harmonization process, the systems measurements are compared against each other in order to verify that the deviation between systems measurements is within acceptable low threshold levels. At the end of the harmonization process all the units provide the same degree of trust while performing pavement inventories at the same surface conditions. After a short description of the testing location and the harmonization methodology, harmonization results (IRI, MPD and crossfall) are illustrated.

### 3.1 Testing location and Methodology

The harmonization process takes place every year in Fredrikstad (Norway) typically in May. The test area is a county road approximately 4 km long and present different surface conditions. Multiple operators drive at approximately 60 km/h over a stretch of road measuring such parameters as roughness, pavement condition, and geometry. Figure 5 shows the unfiltered results of IRI, MPD and crossfall of 6 different harmonized systems driving over the same testing road.

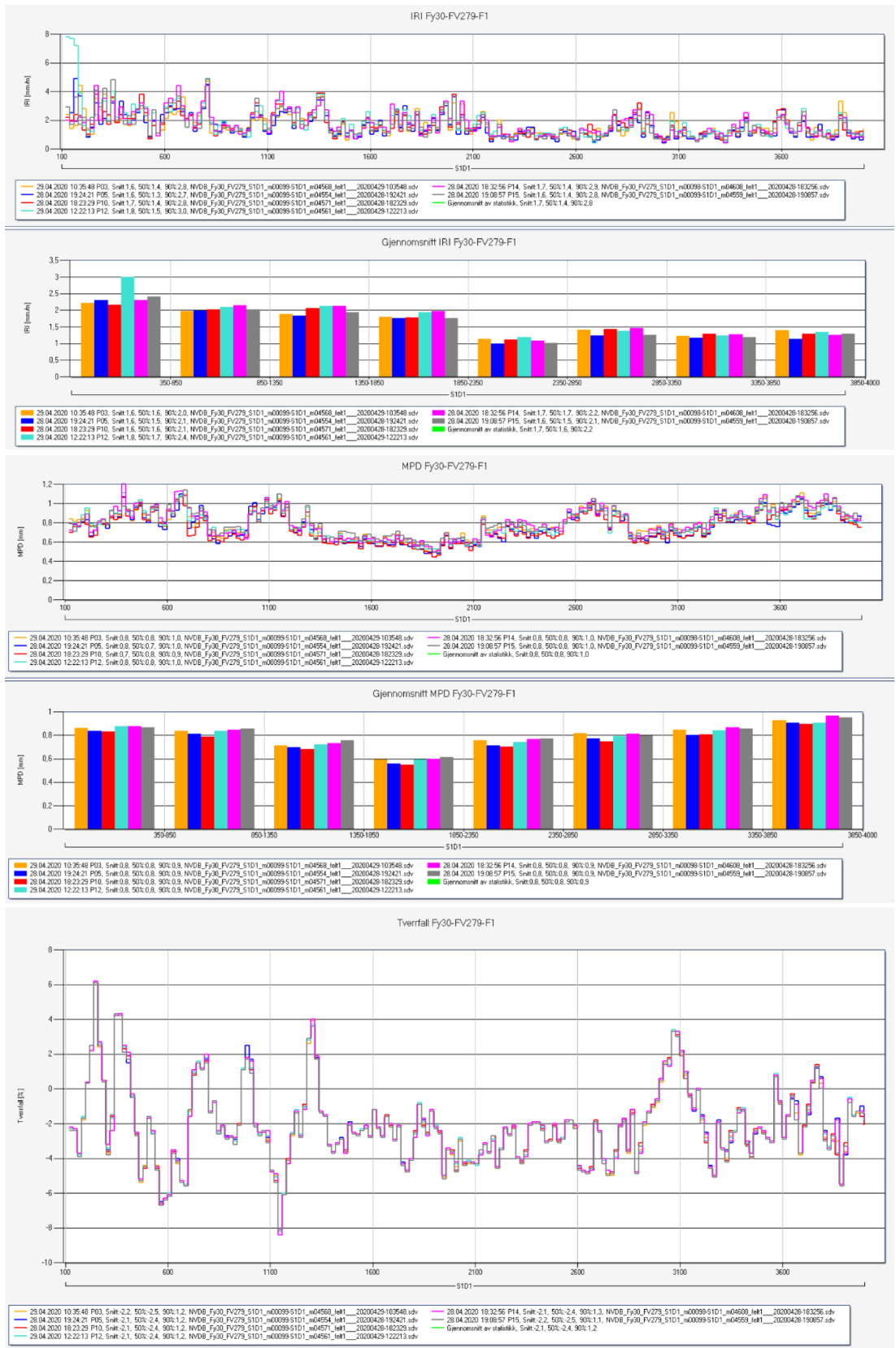

**Figure 5** System results after harmonization process a) IRI; b) MPD; c) crossfall (tverfall).).

Here we illustrate how well the systems correlate after the harmonization process. Figure 6 shows the correlation on the harmonized IRI measurement for different systems. Similarly, Figures 7 and 8 show the harmonized MPD and the harmonized crossfall measurements system comparison.

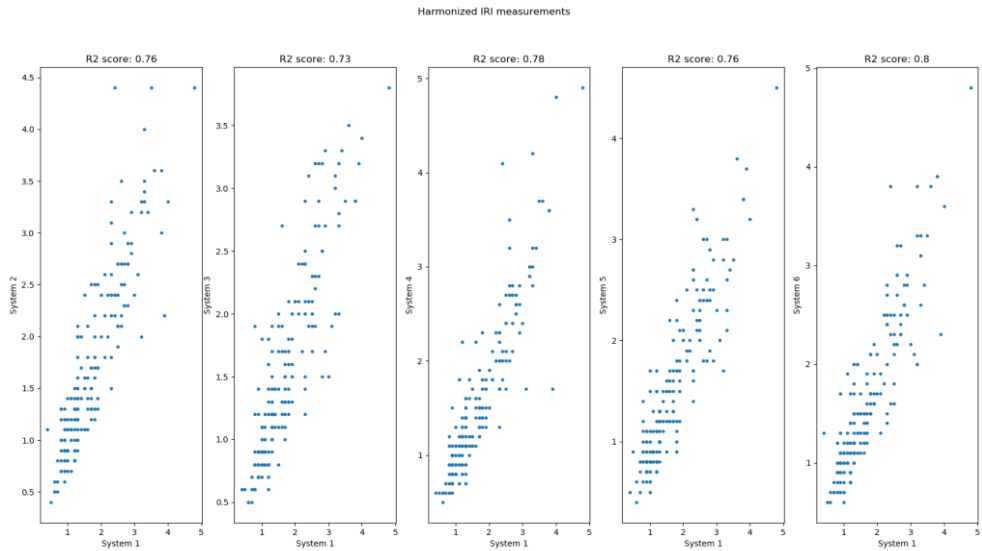

**Figure 6** Harmonized IRI measurement for different system comparison.

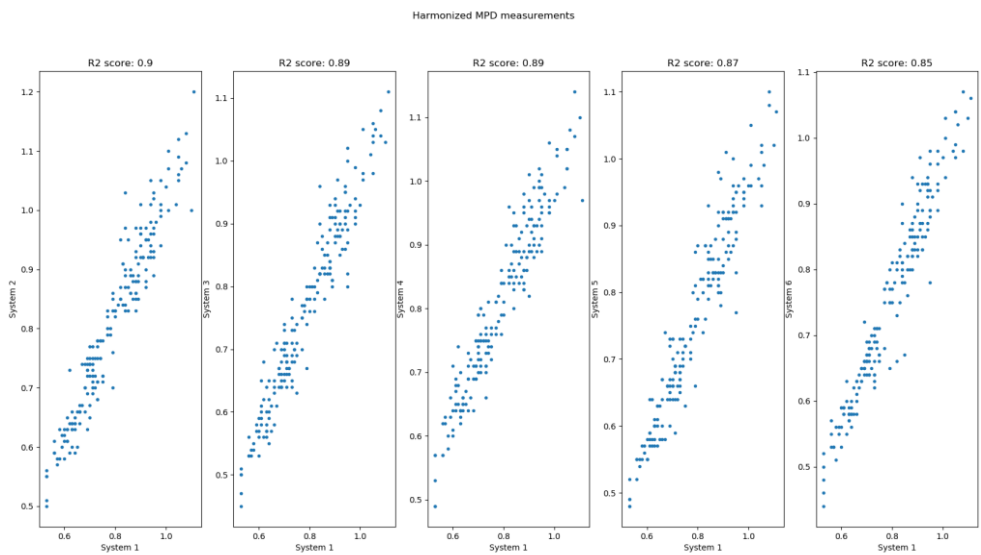

**Figure 7** Harmonized MPD measurement for different system comparison.

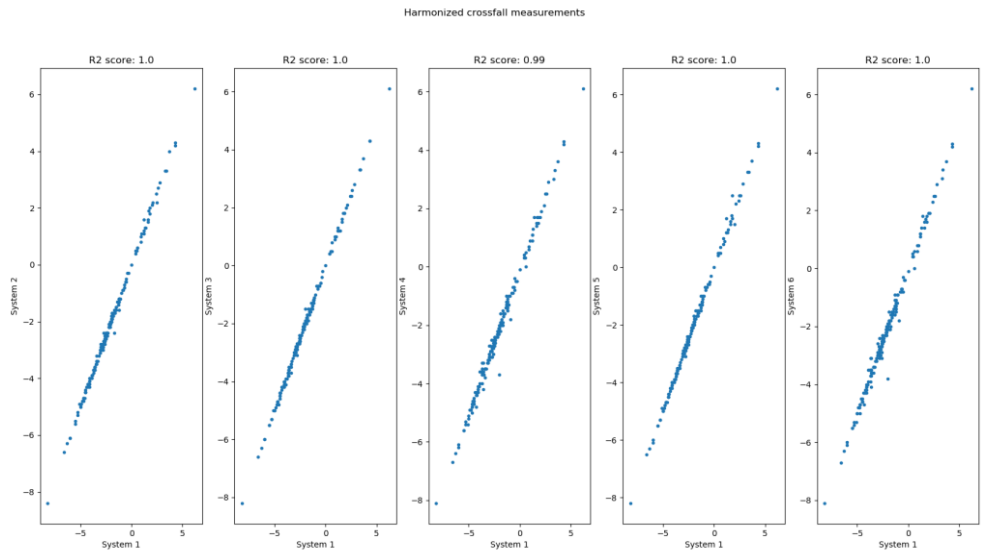

**Figure 8** Harmonized crossfall measurement for different system comparison.

Figure 6 shows for the harmonized IRI measurement an $R^2$ scored ranged from 0.73 to 0.80 while Figure 7 shows for the harmonized MPD an $R^2$ ranged between 0.85 and 0.9. Figure 8 shows for the harmonized crossfall measurements an $R^2$ ranged between 0.99 and 1.

It has to be noted that different operators driving over the testing road in different road conditions is most likely to introduce variance in the measurements, which negatively affect the correlation results. Nevertheless, the results show high positive correlations between the different systems providing confidence in different measurements systems.

## 4 Conclusions

This paper introduced ViaPPS – a 3D Mobile Mapping System from ViaTech AS. The ViaPPS is recognized internationally for its robust, reliable, and accurate georeferenced 3D point cloud solution in applications to pavement inventories. For pavement inventories purposes, ViaPPS is able to detect and geotag according to multiple standards, listed in the reference [9-19], for road defects (cracks, potholes, ravelling, etc) as well as road parameters such as road markings, curve radius, longitudinal and transversal profile, rut depth (area and volume), pavement cross fall, local longitudinal deflection, speed bumps, lane width, vertical height to bridges/crossings or in tunnels, maximum local cross fall in lane and joints. Today the system is adopted worldwide from road practitioners and industrial partners.

# References:

[1] Tao, C. V., & Li, J. (Eds.). (2007). *Advances in mobile mapping technology* (Vol. 4). CRC Press. Burningham, S., & Stankevich, N. (2005).

[2] El-Sheimy, N. (2005). An Overview of Mobile Mapping Systems. Proceedings of FIG Working Week (pp. 16-21).

[3] Harrap, R., & Lato, M. (2010). An overview of LIDAR: collection to application. *NGI publication*, *2*, 1-9.

[4] Guan, H., Li, J., Cao, S., & Yu, Y. (2016). Use of mobile LiDAR in road information inventory: A review. *International Journal of Image and Data Fusion*, *7*(3), 219-242

[5] International Organization for Standardization, 1997. ISO 13473-1: 1997E. Characterization of pavement texture by use of surface profiles – part 1: determination of mean profile depth.

[6] PIARC World Road Association. Report of the Committee on Surface Characteristics. In Proceeding of XVIII World Road Congress, Brussels, Belgium, 13–19 September 1987.

[7] Giesko, T., Zbrowski, A., & Czajka, P. (2007). Laser profilometers for surface inspection and profile measurement. *Problemy Eksploatacji*, 97-108.

[8] Ahmad, N., Ghazilla, R. A. R., Khairi, N. M., & Kasi, V. (2013). Reviews on various inertial measurement unit (IMU) sensor applications. *International Journal of Signal Processing Systems*, *1*(2), 256-262.

[9] AASHTO R86-18, (2019). Collecting Images of Pavement Surfaces for Distress Detection

[10] AASHTO R87-18, (2019). Determining Pavement Deformation Parameters and Cross Slope from Collected Transverse Profiles

[11] AASHTO R88-18, (2019). Collecting the Transverse Pavement Profile

[12] AASHTO E950, (2009). Standard test method for measuring the longitudinal profile of travelled surfaces with an accelerometer established inertial profiling reference

[13] ASTM E1845-09, (2005). Standard Practice for Calculating Pavement Macrotexture Mean Profile Depth

[14] CEN ELEC – EN60825-1, (2014). Safety of Laser Products – Part 1: Equipment Classification and Requirements

[15] EN 13036-8, (2008). Road and airfield surface characteristics. Test methods. Determination of transverse unevenness indices

[16] PIARC R14, (2008). Evaluating the performance of automated pavement cracking measurement equipment

[17] AASHTO PP 37-04, Determination of Interantional Roughness Index(IRI) to Quantify Roughness of Pavements

[18] ISO EN ISO 13473-5, (2009). Characterization of pavement texture by use of surface profiles-Part 5: Determination of megatexture

[19] ISO EN ISO 13473-1, (2019). Characterization of pavement texture by use of surface profiles-Part 1: Determination of Mean Profile Depth

[20] M Velas, M Spanel, Z Materna, A Herout. "Calibration of RGB Camera With Velodyne LiDAR," In WSCG, pp. 1-10, 2014.

[21] F. Vasconcelos, J. P. Barreto and U. Nunes, "A Minimal Solution for the Extrinsic Calibration of a Camera and a Laser-Rangefinder," in IEEE Transactions on Pattern Analysis and Machine Intelligence, vol. 34, no. 11, pp. 2097-2107, Nov. 2012.

[22] J. Heikkila and O. Silven, "A four-step camera calibration procedure with implicit image correction," Proceedings of IEEE Computer Society Conference on Computer Vision and Pattern Recognition, San Juan, Puerto Rico, USA, pp. 1106-1112, 1997.

[23] Lyu, Yecheng and Bai, Lin and Elhousni, Mahdi and Huang, Xinming. "An Interactive LiDAR to Camera Calibration," in IEEE High Performance Extreme Computing Conference (HPEC), 2019

[24] Remondino, F. and C. Fraser. "Digital camera calibration methods: Considerations and comparisons." The International Archives of the Photogrammetry, Remote Sensing and Spatial Information Sciences, pp. 266-272, 2006.

[25] J. Heikkila and O. Silven, "A four-step camera calibration procedure with implicit image correction," in Proceedings of IEEE Computer Society Conference on Computer Vision and Pattern Recognition, San Juan, Puerto Rico, USA, pp. 1106-1112, 1997.

[26] T. Shao-xiong, L. Shan and L. Zong-ming, "Levenberg-Marquardt algorithm based nonlinear optimization of camera calibration for relative measurement," in Proceedings of the 34th Chinese Control Conference (CCC), Hangzhou, pp. 4868-4872, 2015.

[27] Abdel-Aziz, Y. I. & Karara, H. M, "Direct linear transformation into object space coordinates in close-range photogrammetry," in Proc. Symposium on Close-Range Photogrammetry, Urbana, Illinois, p. 1-18, 1971.

[28] K. Levenberg, "A method for the solution of certain problems in least squares". Quarterly of Applied Mathematics, 5, 164–168, 1944.

[29] D. Marquardt, "An algorithm for least-squares estimation of nonlinear parameters." SIAM Journal on Applied Mathematics, 11(2), 431–441, June 1963.